\documentclass[sigconf]{acmart}
\AtBeginDocument{%
  }

\copyrightyear{2025}
\acmYear{2025}
\setcopyright{cc}
\setcctype[4.0]{by}
\acmConference[WWW Companion '25]{Companion Proceedings of the ACM Web Conference 2025}{April 28-May 2, 2025}{Sydney, NSW, Australia}
\acmBooktitle{Companion Proceedings of the ACM Web Conference 2025 (WWW Companion '25), April 28-May 2, 2025, Sydney, NSW, Australia}
\acmDOI{10.1145/3701716.3715246}
\acmISBN{979-8-4007-1331-6/2025/04}

\usepackage[utf8]{inputenc}
\usepackage{microtype}
\usepackage{multirow}
\usepackage{subfigure}
\usepackage[utf8]{inputenc}
\usepackage{graphicx}
\usepackage{amsmath}
\usepackage{algorithm}
\usepackage{algorithmic}
\usepackage{CJKutf8}
\usepackage{balance}
\usepackage[skip=2.5pt]{caption}
\usepackage{array}
\usepackage{booktabs}
\usepackage{threeparttable}
\usepackage{soul}
\usepackage{setspace}
\usepackage{balance}

\begin{document}

\title{LREF: A Novel LLM-based Relevance Framework for E-commerce Search}

\author{Tian Tang}
\authornotemark[1]
\email{tangtian0456@gmail.com}
\affiliation{
  \institution{JD.COM}
  \city{Beijing}
  \country{China}
}

\author{Zhixing Tian}
\authornote{Both authors are corresponding authors.}
\email{tianzhixing2017@gmail.com}
\affiliation{
  \institution{JD.COM}
  \city{Beijing}
  \country{China}
}

\author{Zhenyu Zhu}
\email{zhenzhuyu240@gmail.com}
\affiliation{
  \institution{JD.COM}
  \city{Beijing}
  \country{China}
}
\author{Chenyang Wang}
\email{wangchenyang3@jd.com}
\affiliation{
  \institution{JD.COM}
  \city{Beijing}
  \country{China}
}

\author{Haiqing Hu}
\email{huhaiqing1@jd.com}
\affiliation{
  \institution{JD.COM}
  \city{Beijing}
  \country{China}
}
\author{Guoyu Tang}
\email{tangguoyu@jd.com}
\affiliation{
  \institution{JD.COM}
  \city{Beijing}
  \country{China}
}
\author{Lin Liu}
\email{liulin1@jd.com}
\affiliation{
  \institution{JD.COM}
  \city{Beijing}
  \country{China}
}
\author{Sulong Xu}
\email{xusulong@jd.com}
\affiliation{
  \institution{JD.COM}
  \city{Beijing}
  \country{China}
}

\renewcommand{\shortauthors}{Tian Tang et al.}

\begin{abstract}
Query and product relevance prediction is a critical component for ensuring a smooth user experience in e-commerce search. Traditional studies mainly focus on BERT-based models to assess the semantic relevance between queries and products. However, the discriminative paradigm and limited knowledge capacity of these approaches restrict their ability to comprehend the relevance between queries and products fully. With the rapid advancement of Large Language Models (LLMs), recent research has begun to explore their application to industrial search systems, as LLMs provide extensive world knowledge and flexible optimization for reasoning processes. Nonetheless, directly leveraging LLMs for relevance prediction tasks introduces new challenges, including a high demand for data quality, the necessity for meticulous optimization of reasoning processes, and an optimistic bias that can result in over-recall.

To overcome the above problems, this paper proposes a novel framework called the \textbf{L}LM-based \textbf{RE}levance \textbf{F}ramework (\textbf{LREF}) aimed at enhancing e-commerce search relevance. The framework comprises three main stages: supervised fine-tuning (SFT) with Data Selection, Multiple Chain of Thought (Multi-CoT) tuning, and Direct Preference Optimization (DPO) for de-biasing. We evaluate the performance of the framework through a series of offline experiments on large-scale real-world datasets, as well as online A/B testing. The results indicate significant improvements in both offline and online metrics. Ultimately, the model was deployed in a well-known e-commerce application, yielding substantial commercial benefits.

\end{abstract}




\maketitle

\section{Introduction}
E-commerce has become a vital part of daily life, revolutionizing our shopping habits. Platforms like Amazon, Taobao, and JD offer an extensive range of products. The product search system, comprising retrieval and ranking modules, effectively connects user demand with available products\cite{yuan2023multi}. The heart of this system is search relevance, a crucial factor for enhancing user experience, as users expect results that precisely meet their needs\cite{liu2022knowledge}.

Most previous methods employ BERT-like pre-trained models for search relevance tasks. These models provide a foundational level of context encoding and language understanding, which ensures basic relevance. However, despite their pre-training, these models possess limited world knowledge due to their constrained parameter scale. This limitation hinders their ability to fully comprehend the diverse range of e-commerce products and user expressions. 
Additionally, their training methods face constraints. Due to the encoder architecture, BERT-like models inherently adopt a discriminative pattern for relevant tasks. This pattern relies on end-to-end data-driven training, making it challenging to optimize the internal reasoning processes\cite{wang2024utilizing}.
Based on these observations and insights, We choose to base search relevance on large language models (LLMs) instead. Through the pre-training phase, LLMs acquire extensive world knowledge, which is crucial for understanding the wide variety of products and user queries. Unlike discriminative models, LLMs possess strong reasoning capabilities and employ a generative paradigm to solve application problems. This paradigm allows for a more controllable and optimizable decision-making process regarding the relevance of a product to a user's query, which significantly enhances training flexibility\cite{chang2024survey}.

Due to LLMs' inherent characteristics, applying them to e-commerce search relevance presents three major challenges. (1) Quality Rather Than Quantity Data\cite{li2023quantity}: LLMs undergo extensive pre-training on large datasets and unify downstream tasks under a generative paradigm. As a result, they already possess an initial capability for text-matching tasks, similar to search relevance tasks, even before specific training. Therefore, when further optimizing for relevant tasks, the quality of the data becomes more critical than its quantity. This necessitates targeted Data Selection from a task perspective.
(2) Task-Specific Intermediate Reasoning: As previously mentioned, the generative paradigm allows LLMs to further optimize the decision-making process for downstream tasks. However, this process requires adherence to specific domain task rules. It is challenging to ensure that the model learns and follows these finely-tuned business rules.
(3) Optimistic Bias: When applied to relevant tasks, LLMs tend to make more lenient and optimistic judgments, possibly due to their value alignment processes. This characteristic can lead to over-recall, resulting in the mis-exposure of irrelevant products, which negatively impacts user experience\cite{eigner2024determinants}.

To address these challenges, we propose the \textbf{L}LM-based \textbf{RE}levance \textbf{F}ramework (\textbf{LREF}), which comprises three key components: (1) Data Selection: We perform targeted selection of annotated data by considering the characteristics of relevance tasks, the real-world distribution of online queries, and feedback from the LLM model. Our selection strategy enables the model to achieve better performance with less training data. (2) Multi-CoT Tuning: we guide the LLM to think according to relevant task rules, analyzing complex semantic relationships within query-product pairs and reflecting on incorrect reasoning. This approach positions LLMs as relevance experts, capable of accurately reasoning about user intents.\cite{wei2022chain} (3) DPO De-biasing: To address the issue of excessive recall due to the LLM's over-optimistic judgments, we model this as a preference optimization problem. By introducing DPO methods, we reinforce its learning of the ambiguous query-product pairs and guide LLMs to make objective and unbiased judgments in borderline cases.\cite{rafailov2024direct}
We evaluate the performance of LREF with large-scale offline datasets, as well as online A/B testing. The results demonstrate notable enhancements in both offline and online metrics.  Ultimately, the model was deployed in a well-known e-commerce application, delivering considerable commercial advantages.
The contributions of this paper can be summarized as follows:
\begin{itemize}
    \item We propose an innovative LLM-based framework for product search relevance, named LREF, which demonstrates advantages over previous BERT-like approaches.
    \item We introduce a novel and practical Data Selection approach for the LLM-based relevance method. 
    \item We design the Multi-CoT Tuning strategy to optimize the internal reasoning processes for search relevance.
    \item The DPO strategy is applied to reinforce the LLM's performance in boarding query-product pairs and de-bias LLM's over-recall problem.
\end{itemize}
In this section, we will discuss the previous studies related to our work: the relevant language models and the supervised fine-tuning of large language models.

\subsection{Search Relevance}
There is plenty of literature about search relevance technology, and some of the older works focus on Keyword Matching, in their framework, the search engine relies on TF-IDF(Term Frequency-Inverse Document Frequency) \cite{aizawa2003information}. As the development of Natural Language Processing (NLP), the synonyms, and related terms are to be considered by using the Word2Vec\cite{mikolov2013distributed} and GLove\cite{pennington2014glove}. However, these method has the limitation of capturing user intents and analyzing contextual semantics. Over time, Transformer-based models\cite{vaswani2017attention} have revolutionized the field, the relevance problems not only solved by language models but also by incorporating user feedback into the learning process\cite{yan2018beyond}. Furthermore,graph-based relevance problem-solving is also presented in improving search relevance.\cite{choudhary2024interpretable}. With the rapid advancement of technology, the model size and complexity are increasing, and approaches like DeBERTa\cite{he2020deberta} and extended BERT pretraining demonstrate the importance of larger-scale models. However, the constrained knowledge capacity of these approaches limits their ability to fully understand users' intents and complex e-commerce scenarios\cite{yuan2024semi}. Recently, the emergence of LLMs has dramatically expanded model parameters and has a strong performance in natural language tasks\cite{touvron2023llama}. It is essential for industry frameworks to quickly leverage the advantages of the latest models. In our framework, unlike most of the industry end-to-end data fine-tuning\cite{mehrdad2024large}, we propose a novel LLM-based framework for product search relevance.
\begin{figure*}[!htbp]
    \centering
    \includegraphics[scale=0.035]{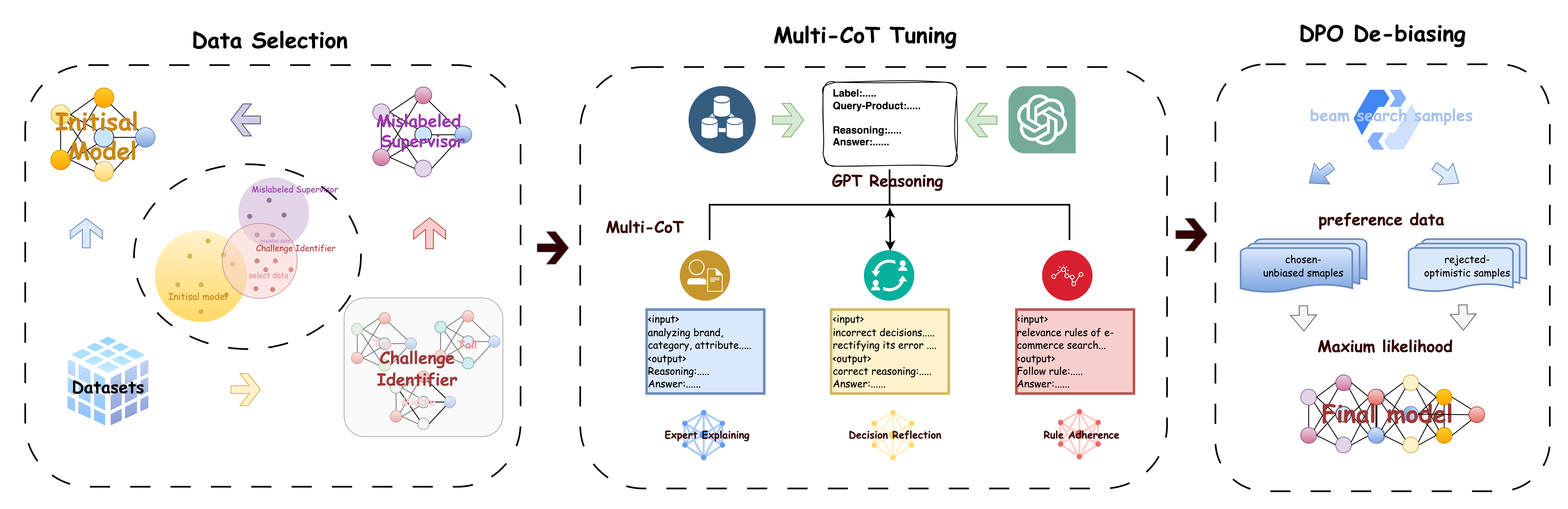}
    \caption{LREF: a novel LLM-based Relevance Framework for Product Search}
    \label{model_structure}
\end{figure*}

\subsection{Large Language Model}
With the advent of large language models (LLMs), OpenAI’s GPT-3\cite{floridi2020gpt} introduced few-shot and zero-shot capabilities which enhance model capabilities in domain-specific tasks without requiring extensive task-specific data\cite{gao2023llama}. However, supervised fine-tuning (SFT) plays a crucial role in achieving optimal performance in targeted applications by adjusting the model's parameters to improve performance for tasks.\cite{naveed2023comprehensive}. Among recent SFT methodologies, Chain-of-Thought (CoT) \cite{wei2022chain} prompting has been influential in reasoning-intensive tasks. For instance, Self-Consistency CoT \cite{wang2022self} improves logical consistency and accuracy in complex problem-solving scenarios. Additionally, the quality of training data has emerged as a decisive factor in the finetuning process\cite{xia2024less}, the high-quality datasets dramatically enhance the task-solving capabilities of large language models and allow a more effective training process\cite{liu2024take,cao2023instruction}. Moreover, supervised fine-tuning incorporates Reinforcement Learning from Human Feedback (RLHF), enhancing the model's ability to understand user intents\cite{kaufmann2023survey,ethayarajh2024kto}. To address the search relevance domain-specific tasks, our three-stage framework includes Data Selection, Chain of Thought (CoT) tuning\cite{wei2022chain}, and direct preference optimization (DPO) \cite{rafailov2024direct} for de-biasing. This combined approach allows the models to improve the relevance of domain-specific task ability while maintaining efficiency.

\section{Methods}


The task is product search relevance, which can be formulated as a classification problem. Drawing from a real-world e-commerce application, we have developed the ESMTR classification schema, which includes the categories: Exact, Significant, Marginal, Trivial, and Irrelevant. For a given query-product pair, the model is tasked with predicting the appropriate label.


\subsection{Methods Overview}
In this part, we introduce the proposed LREF method. As shown in Figure \ref{model_structure}, LREF is mainly composed of three stages: As LLMs normally require data of high quality, we first adopt a novel and practical Data Selection approach for Supervised fine-tuning (SFT). Subsequently, the Multi-CoT tuning strategy is applied to guide the progress of the LLM reasoning for relevance judgment. Finally, faced with the optimistic bias problem, we tune the model with Direct preference optimization (DPO) for better performance on the marginal cases.

Specifically, the select data module is used to select reliable and influential datasets for the model. The Multi-CoT module bolsters model reasoning ability in search relevance environment and the DPO De-biasing model training model to deal with the boundary problem effectively.

\subsection{SFT with Data Selection}
Holding initial strong capability for text-matching tasks, LLM is sensitive to data quality\cite{wang2024survey}rather than relying on vast data, during SFT, we focus on selecting influential data from large human-labeled data. Specifically, we aim to select challenging but not noisy relevant examples. To achieve this, we have specifically designed three auxiliary models: Initial Model, Challenge Identifier, and Mislabeled Supervisor. Those three models are initialed from the same LLM, but trained with different data.

The \textbf{Initial Model} (IM), built upon an open-source LLM, is fine-tuned using random samples from human-labeled datasets, where common query-product pairs are prevalent. As a result, it performs well on typical examples. In contrast, the \textbf{Challenge Identifier} (CI) is designed to recognize more challenging examples and is trained on a more balanced and diverse set of query-product samples. Specifically, we utilize an e-commerce query online feature to categorize these pairs into three types based on the exploring distribution: top (common), middle, and long tail. We then sample these pairs uniformly to ensure that the middle and long-tail pairs have a significant impact. Specifically, the Challenge Identifier identifies as follows:
\begin{equation}
S_{\text {seed }}=\{x \in D \mid C I(x) \text { is correct }\}
 \end{equation}
\begin{equation}
S_{\text {challenging }}=\left\{x \in S_{\text {seed }} \mid I M(x) \neq L(x)\right\}
\end{equation}
the Dataset $D$ is the full training dataset, $CI(x)$ is the prediction of the Challenge Identifier, the Labels $L(x)$ is the set of human-labeled classes and the $IM(x)$ denote to the prediction of the Initial Model for $x$.So firstly we select a seed sample with CI then filter out challenging sample samples with $IM$.

In the next step, the \textbf{Mislabeled Supervisor} (MS) is designed to identify and filter out potentially mislabeled and ambiguous annotations within datasets. We randomly select samples from the annotation datasets and employ GPT to ascertain the most confounding classes for these samples. For instance, a typical prompt might be: "In an e-commerce search environment, given the query 'iPhone' and a product described as 'Apple 20W USB-C Phone Charger Original Fast Charger for iPhone,' the current relevance class is 'Trivial.' What would be the most confounding alternative relevance class?" In this context, GPT might suggest 'Marginal.' These ambiguous labels are subsequently used to fine-tune the Mislabeled Supervisor, enhancing its ability to predict the most confounding labels. This methodology assists the supervisor in identifying and filtering out annotations that may be erroneously labeled by humans, thereby purifying the training datasets from noise. So the filter out samples where the mislabeled supervisor suggests possible labeling noise. The process is defined as follows:
\begin{equation}
S_{\text {selection }}=\left\{x \in S_{\text {challenging }} \mid M S(x) \neq L(x)\right\}
\end{equation}

In the Data Selection process, we employ these three models to evaluate the entire training dataset. We begin by selecting the samples correctly predicted by the Challenge Identifier as seed samples. From these, we filter out the simple samples that are also correctly predicted by the Initial Model. Next, we remove potentially mislabeled samples, which are those where the Mislabeled Supervisor's predicted labels match the human annotations. This process yields a set of noise-free, valuable challenging samples. The final selection set $S$ is defined as follows: 
\begin{equation}
S_{\text {selection }}=\{x \in D \mid C I(x) \text { is correct, } I M(x) \neq L(x), M S(x) \neq L(x)\}
\end{equation}
Finally, we use these samples to further SFT the Initial Model by a Language Model (LM) loss:
 \begin{equation}
 \resizebox{.64\hsize}{!}{$\mathcal{L}_{lm} = -\sum_{t=1}^{T} \log P(w_{t+1} \mid w_1, w_2, \ldots, w_t)$}
 \end{equation}
resulting in a model that performs better than one trained on the entire annotated dataset. The specific results of this comparison will be detailed in the experimental section.

\subsection{Multi-CoT Tuning}

Unlike traditional BERT-like models that primarily depend on end-to-end learning, large language models (LLMs) leveraging the generative paradigm can improve their internal decision-making capabilities for downstream tasks\cite{zhang2022automatic}. Nonetheless, achieving this requires strict compliance with domain-specific task guidelines. Ensuring that the model handles the complex e-commerce domain and adheres to these meticulously crafted business rules remains a significant challenge. For product search relevance, we designed the Multiple Chain of Thought Tuning (Multi-CoT Tuning) strategy to address the problem. We introduce three types of CoT: Expert Explaining (EE), Rule Adherence (RA), and Decision Reflection (DR).

\textbf{Expert Explaining} Chain of Thought (EE-CoT) is introduced to offer domain-specific analysis from multiple perspectives. \cite{chu2023survey}Given a training data point consisting of a user query, product title, and relevance label, represented as (query, title, label), we provide this triplet, including the label, to the GPT model. The model is then tasked with analyzing the query and product from multiple dimensions—such as brand, category, attribute, and keywords—using the relevance label as a guide. This analysis results in an explanation of the EE-CoT. Subsequently, we construct a quadruplet (query, title, label, EE-CoT) and input the (query, title) pair into our LLM, training it to generate the EE-CoT and label. This constitutes a form of thought-guided training. 

Furthermore, we developed the \textbf{Rule Adherence} Chain-of-Thought (RA-CoT) to guide the LLM in adhering to the relevance rules of a real e-commerce search system. This rule defines query-product relevance through two primary components: product relevance and modifier relevance. Product relevance assesses whether the item meets the basic product type and functional requirements, such as a Samsung S24 being relevant at the product level to a query for "iPhone 15" based on product type and functionality. Modifier relevance focuses on specific descriptive attributes like brand, model, and features; in the previous example, the brands Apple and Samsung do not match, indicating a lack of modifier relevance.
Based on the alignment across these two dimensions, a final five-tier relevance judgment is made, categorized as Exact, Significant, Marginal, Trivial, or Irrelevant. We input this complex judgment rule alongside the original sample to form a quadruplet (Rule, query, title, label), which is provided to the GPT model. The model uses the rule to generate a corresponding reasoning process as the RA-CoT, resulting in a quintuplet (Rule, query, title, label, RA-CoT). Subsequently, we input the (query, title, rule) into the model, training it to generate both the RA-CoT and the label.

In addition, to forward explanation and reasoning, we implemented a \textbf{Decision Reflection} Chain of Thought (DR-CoT)\cite{renze2024self}. This approach involves leveraging incorrect decisions made by the base LLM to guide the model in rectifying its errors. We provide the GPT model with these incorrect predictions, which are combined with the original sample information to form a tuple (incorrect decision, query, title, label). The model is then tasked with generating the correct reasoning path and identifying the flaws in the initial predictions. This corrective reasoning, referred to as DR-CoT, is subsequently fed back into the base LLM. Specifically, the LLM is later provided with (query, title, incorrect decision) as input, and it is trained to output both the DR-CoT and the correct label.

\subsection{DPO De-biasing}
After performing Supervised Fine-Tuning (SFT) with Data Selection and Multi-CoT tuning, further analysis of the prediction samples revealed the emergence of an optimistic bias in the LLM model\cite{villaflor2022addressing}.In relevance tasks, when faced with uncertainty, the LLM tends to classify a marginal query-product pair as significant, thus erroneously deeming it relevant. Specifically, after two stages of training, we found that the model still incorrectly predicted 9\% of the training data. Of these incorrect predictions, 70\% were misclassified as significant to the query-product pair, resulting in a state of over-recall. However, when we increase the beam size during prediction, we find that 80\% of the previously overestimated cases are corrected in the second position of the output. Based on these observations and analyses, we modeled this issue as a preference alignment problem, aiming to shift the LLM's approach from overly optimistic to objective and cautious in relevant tasks. To achieve this adjustment, We employed the Direct Preference Optimization (DPO) method:
\begin{equation}
\resizebox{.80\hsize}{!}{$\mathcal{L}_{\text{DPO}} = -\mathbb{E}{(x, y^+, y^-)} \left[ \log \sigma(f_\theta(x, y^+) - f_\theta(x, y^-)) \right]$}
\end{equation}
where $f_\theta(x, y)$ is the LLM with parameters $\theta$, $(x, y^+, y^-)$ represents an input and its corresponding preference pair. $ y^+$ is the chosen output and $y^-$ is the rejected one.
Specifically, we selected samples from the training data where the LLM, trained by the previous two stages, made incorrect predictions, but the correct answer appeared within the top k positions during the beam search process\cite{wiseman2016sequence}. we choose the incorrect answer as the rejected data $ y^-$, while the correct answer from beam search is chosen data $ y^+$. This approach reinforces the model's learning of the preference order among relevant data classes and guides it in making objective and unbiased predictions, particularly in cases of weak relevance.

\begin{table*}[!htbp]
  \caption{
        Offline Evaluation Results 
  }
  \centering
  \label{tab:experiment_overall}
  \setlength{\tabcolsep}{4mm}{
      \begin{tabular}{c|ccc|ccc}
                \toprule
                \multirow{2}{*}{\textbf{Models}}  & 
                \multicolumn{3}{c|}{\textbf{ESMTR classification}} & \multicolumn{3}{c}{\textbf{ Relevant classification}} \\
                            
                &\textbf{Macro F1} &\textbf{Weighted F1} &\textbf{Accuracy}
                &\textbf{Precision} &\textbf{Recall} &\textbf{F1}\\
                \midrule
                \midrule
                BERT      & 54.09 & 63.71 & 63.86 & 81.21 & 80.72 &80.96 \\
                DeBERTa   & 53.25 & 64.78 & 64.78 & 80.10 & 85.10 &82.52  \\   
                \midrule 
                LLM Base            & 44.86  & 59.63  & 62.80 & 81.76 & 78.61 &80.16 \\
                \midrule 
                LREF (DataSelect)        & 53.00 & 65.07  & 65.74  & 83.68 & 82.46 &83.07\\
                LREF (DataSelect\&Multi-CoT)       & \textbf{56.21}  &  66.03 & 66.23 & 82.72 & \textbf{86.12} & 84.39\\
                LREF (DataSelect\&Multi-CoT\&DPO)      & 55.90 &  \textbf{66.91} & \textbf{67.08}  & \textbf{84.12} & 85.82& \textbf{84.96}\\
                \bottomrule
        \end{tabular}
    }
\end{table*}

\begin{table}
  \caption{Distribution of Test Datasets}
  \label{tab:datset}
  \setlength{\tabcolsep}{5mm}{
      \begin{tabular}{ccl}
        \toprule
        Relation & Number&Proportion\\
        \midrule
        Irrelevant & 3135& 0.95 \\
        Trivial & 90057& 27.29 \\
        Marginal & 61809& 18.73 \\
        Significant & 135564& 41.08 \\
        Exact & 39435& 11.95 \\
        \midrule
        total & 330000& 100 \\
        \bottomrule
      \end{tabular}
   }
\end{table}

\section{Experiment}
This section provides a detailed exploration of the offline and online experiments. We first introduce the dataset composition and class distribution of our test datasets. Next, analyze the experiment results with several strong baseline models. Following this, we will show the Data Selection experiences at each selection stage, highlighting their impact. We then compare the different types of COTs and the DPO improvements of the LLM model. Subsequently, we exhibit the online experiences of the model on the JD ab test platform and analyze its effect on user engagement and performance metrics.

\subsection{Datasets and Metrics}
\label{sec:Dataset}
The test datasets are directly collected from the JD application online users' click logs, with human annotators providing the labels. The datasets consist of 33,000 queries-product pairs and the statistics of the datasets are listed in Table~\ref{tab:datset}. Compared with the online query-product distribution, the test datasets involve a higher proportion of Significant, Marginal, and Trivial classes to evaluate the model's capability in addressing boundary cases. The training datasets are approximately from the same distribution as the test datasets, comprising a total of 5,000,000 query-product pairs. 

We evaluate our framework from two perspectives: the ESMTR classification and the irrelevant classification. the ESMTR classifications are divided into five classes and applied for the online product show order. The relevant classification is used to enhance the user experiences in online marketing. We evaluate our framework by weighted F1 scores, macro F1 scores, and accuracy (ACC) metrics. Each evaluation metrics are applied to validate the framework's effectiveness in real-world applications.

\subsection{Baselines}
For the baselines, we compare our models with several strong baseline models including LLAMA-2-7B~\cite{gao2023llama}, BERT~\cite{devlin2018bert}, and DeBERTa~\cite{he2020deberta}. The evaluation focuses on assessing the fundamental text-processing capabilities of the LLM and determining whether our framework outperforms other benchmarks. 
\begin{itemize}
    \item \textbf{BERT}: transformer-based language model designed for understanding the context of words in natural language. We train the model on the full train set.
    \item \textbf{DeBERTa}: Introduced by Microsoft in 2020, it improves upon BERT by addressing limitations in capturing word order and position information. We train the model on the full train set.
    \item \textbf{LLM Base}: We fine-tune the open-source LLM, LLaMA-2 7B, on a full train set and then obtain the LLM Base model.
    \item \textbf{LREF}: the proposed LLM-based Relevance Framework (\textbf{LREF}), initialed from LLaMA-2 7B, and optimized by Data select SFT, Multi-CoT Tuning and DPO de-biasing strategies. 
\end{itemize}

\subsection{Implementation Detail}
We employed the 7B LLM to build LREF. All training process on 8 H-800 GPUs, with the batch sizes to 16. The warm-up ratio is 0.2 and the deep-speed stage is 1. Furthermore, we use the Adamw optimizer with the learning rate  $2e^{-5}$. The max length of the model training is 500 and the epoch is 8. To overcome the overfitting,  we save the best checkpoints in the end and use validation datasets as a split of 10 percent of the training datasets.  As for Dpo training the $\alpha$ is set to 0.65. The epoch is to be set as 2 to prevent overfitting. For the inference part, we utilize vLLM with the following settings with the temperature set as 0 for deterministic outputs. 

\subsection{Offline Evaluation}

\subsubsection{\textbf{Overall Performance}} As shown in Table~\ref{tab:experiment_overall}, our proposed LREF method achieves significant improvements compared to the baselines. 
Furthermore, taking LLM Base as a reference point, we observe that both BERT and DeBERTa outperform it. This indicates that the straightforward application of LLMs to relevant tasks, specifically by fine-tuning LLMs with all annotated data (SFT), does not inherently surpass the performance of classical BERT-like discriminative models.
Only after we address the need for high-quality data in SFT, optimize the internal reasoning steps, and mitigate optimistic bias, our proposed LREF, which also leverages LLMs, achieve optimal performance.

\begin{table*}[!htbp]
  \caption{
        The Ablation Study of Data Selection SFT
  }
  \centering
  \label{tab:Ablation_data}
  \setlength{\tabcolsep}{5mm}{
      \begin{tabular}{c|ccc|ccc}
                \toprule
                \multirow{2}{*}{\textbf{Models}}  & 
                \multicolumn{3}{c|}{\textbf{ESMTR classification}} & \multicolumn{3}{c}{\textbf{Relevant classification}} \\                
                &\textbf{Macro F1} &\textbf{Weighted F1} &\textbf{Accuracy}
                &\textbf{Precision} &\textbf{Recall} &\textbf{F1}\\
                \midrule
                Full Data            & 44.86  & 59.63  & 62.80 & 81.76 & 78.61 &80.16 \\
                \midrule
                DataSelect (IM)      & 46.29  & 59.89 & 60.10 & 78.74 & 83.09 &80.85 \\
                DataSelect (IM\&CI)   & 52.81  & 63.91 & 63.88 & 83.34 & 82.27 &82.81  \\   
                DataSelect (IM\&CI\&MS)    & 53.00 & 65.07  & 65.74  & 83.68 & 82.46 &83.07 \\
                \bottomrule
        \end{tabular}
    }
\end{table*}

\begin{table*}[!htbp]
  \caption{
        The Ablation Study of Multi-CoT Tuning
  }
  \centering
  \label{tab:Ablation_cot}
  \setlength{\tabcolsep}{5mm}{
      \begin{tabular}{c|ccc|ccc}
                \toprule
                \multirow{2}{*}{\textbf{Models}}  & 
                \multicolumn{3}{c|}{\textbf{ESMTR classification}} & \multicolumn{3}{c}{\textbf{Relevant classification}} \\                
                &\textbf{Macro F1} &\textbf{Weighted F1} &\textbf{Accuracy}
                &\textbf{Precision} &\textbf{Recall} &\textbf{F1}\\
                \midrule
                \midrule
                w/o Multi-CoT         & 53.00 & 65.07  & 65.74  & 83.68 & 82.46 &83.07 \\
                \midrule
                Multi-CoT (EE)      & 54.30 & 64.58& 64.38 & 83.02 &85.01& 84.00\\ 
                Multi-CoT (EE\&RA) & 55.40 & 65.27& 65.25   & 83.27&84.24 & 83.75\\
                Multi-CoT (EE\&RA\&DR)   & 56.21  &  66.03 & 66.23 & 82.72 &86.12 & 84.39\\
                \bottomrule
        \end{tabular}
    }
\end{table*}

\begin{table*}[!htbp]
  \caption{
        The Analysis for DPO-biasing
  }
  \centering
  \label{tab:analysis_dpo}
  \setlength{\tabcolsep}{1.8mm}{
      \begin{tabular}{c|ccc|ccc|ccc|ccc}
                \toprule
                \textbf{Class}
                & \multicolumn{3}{c|}{\textbf{LLM Base}}  
                & \multicolumn{3}{c|}{\textbf{LREF}} 
                & \multicolumn{3}{c|}{\textbf{BERTmodel}} 
                & \multicolumn{3}{c}{\textbf{DeBERTamodel}} \\
                
                &\textbf{Precision} &\textbf{Recall} &\textbf{F1}
                &\textbf{Precision} &\textbf{Recall} &\textbf{F1}
                &\textbf{Precision} &\textbf{Recall} &\textbf{F1}
                &\textbf{Precision} &\textbf{Recall} &\textbf{F1}\\
                \midrule
                \midrule           
                Marginal         & 66.88&50.13&57.31& 69.02&64.12&66.48  & 62.86 & 62.30 & 62.58 & 63.88 & 63.73 & 63.80 \\
                Significant     &61.39&73.52&66.91& 66.91&71.18&68.98 & 64.26 & 64.63 & 64.45 & 63.82 & 66.94 & 65.34\\
                \bottomrule
        \end{tabular}
    }
\end{table*}

\subsubsection{\textbf{Effectiveness of Data Selection in SFT}} In  Table~\ref{tab:experiment_overall}, LREF (DataSelect) is the model that SFT with of Data Selection method. Specifically, through the proposed Data Selection method, we obtain about 0.5m noise-free, valuable challenging samples from the train set (about 10\% selected), and sequentially fine-tune the Initial Model, which is already trained on a random 150,000 samples. As shown in Table~\ref{tab:experiment_overall}, the comparison between LLM Base and LREF(DataSelect) demonstrates the effectiveness of our Data Selection method for LLM applying to relevant tasks. 

Table~\ref{tab:Ablation_data} shows the further ablation study of Data Selection in SFT. DataSelect(IM) is the Initial Model(IM) which is built on the same open-source LLM with LLM Base, and trained on random 0.15m samples. DataSelect(IM\&CI) is the model trained from the Initial Model on 0.56m challenging but noisy samples selected by CI and IM. DataSelect(IM\&CI\&MS) represents the model trained from the Initial Model on 0.5m challenging and noise-free samples selected by IM, CI, and MS. 
The results in Table ~\ref{tab:Ablation_data} show LLM Base, the LLM trained on the full train set, only matches the performance of the Initial Model trained on a randomly selected subset of data. However, with the introduction of CI and MS, the data is progressively refined, and the performance of LLM SFT is gradually improved. Ultimately, in the SFT stage, LREF significantly surpasses the performance of the LLM trained on the full dataset by using only a portion of high-quality data.

\subsubsection{\textbf{Effectiveness of Multi-CoT Tuning}} As shown in Table~\ref{tab:experiment_overall}, LREF (DataSelect\&Multi-CoT) reports the performance of Multi-CoT Tuning after SFT with Data Selection. It outperforms LREF (DataSelect) significantly, which proves the effectiveness of Multi-CoT tuning which focuses on optimizing the internal reasoning progress of LLM for relevant tasks. 

Table~\ref{tab:Ablation_data} shows the further ablation study of Multi-CoT Tuning. In this table, w/o Multi-CoT means LLM fine-tuned by the selected data, the first stage of LREF. The Multi-CoT (EE) is the LLM trained subsequently from the first stage on the Expert Explaining (EE) Chain of Thought (CoT) data. Multi-CoT (EE\&RA) and Multi-CoT (EE\&RA\&DR) indicate the successive incorporation of Rule Adherence (RA) CoT and Decision Reflection (DR) CoT into the training process, respectively. With the guidance of Expert Explaining, the LLM became more adept at understanding e-commerce search scenarios, leading to significant improvements in the Relevant classification metric, which directly impacts user experience. Furthermore, the incorporation of Rule Adherence enabled the LLM to follow predefined standards during the reasoning process, enhancing overall performance. Finally, with Decision Reflection, the model engaged in targeted reflection on erroneous examples, elevating the effectiveness of the Multi-CoT strategy to a notably higher level.

\subsubsection{\textbf{Effectiveness of DPO De-biasing}}
Through the experiments, we observed that when applied to relevance tasks, LLMs tend to make lenient and optimistic judgments. Specifically, compared to BERT-like baselines, LLMs are more prone to misclassifying Marginal cases as Significant, adopting a more optimistic stance in evaluating the relevance between products and queries. As shown in Table~\ref{tab:analysis_dpo}, the LLM Base exhibits lower recall for the Marginal class and higher precision for the Significant class. Consequently, our LREF approach employs DPO to mitigate this optimistic bias. As indicated by the data in the table~\ref{tab:experiment_overall}, incorporating DPO-debiasing, in addition to Data Selection SFT and Multi-CoT Tuning, further enhances the overall performance of our proposed LREF. Moreover, when examining the precision and recall across different categories, we found significant improvements in Marginal recall and Significant precision. This indicates that the corresponding relevance methods can better control the mis-exposure of non-significantly relevant samples.

\subsection{Online Evaluation}
\subsubsection{\textbf{Online Deployment}}
To reduce the response latency of online deployment, we leverage the data distillation method for transferring LLM ability in online deployment, we employ LLM to annotate approximately 2 billion datasets and then transfer the knowledge to a BERT-based model for efficient online serving. Additionally, the relevance classification is sent to the rank module, which organizes product orders within each relevance class. This means that the relevance prediction results will influence the product display order in the JD apps. For these irreverence products it will be deleted in the ranking stage and not be displayed under the query search pages. This ensures that irrelevant items are filtered out in the ranking pipeline, improving the user experience and online efficiency.

\subsubsection{\textbf{Online Performance}}
To evaluate our framework, we deployed it on the JD A/B test platform. We randomly select a 20\% traffic group as the test group to employ our LREF approach. Another 20\% served as the base group which utilized the previous BERT-based model. For the fair comprising, we continuously monitored experimental performance on the platform. To avoid the effects of traffic fluctuations, we set at least a 7-day testing period.
For online evaluation, the following business metrics were used:
\begin{equation}
\begin{split}
    & UV value = \frac{GMV}{UV}    \,,  \\
    & UCVR = \frac{Orderlines}{UV}    \,,  \\
    & UCTR = \frac{Clicks}{UV}    \,,  \\
\end{split}
\end{equation}
To be specific, the UCVR is the conversion rate of users and UCTR is the click rate of users. Furthermore, Relevance Satisfaction is also used as a key metric, it measures how the displayed products match user intent based on the query. Expert annotators measure the relevance satisfaction, they determine the relevance satisfaction by reviewing each query and evaluating the displayed products case by case. Around 20,000 cases are annotated to generate relevance satisfaction in test and base groups.
\begin{table}[!htbp]
  \centering
  \caption{Online improvements of the LREF.}
  \label{online_performance_uv_ucvr}
  \setlength{\tabcolsep}{1.5mm}{
      \begin{tabular}{c|cccc}
            \toprule
            Models &\textbf{UV value} &\textbf{UCVR}  & \textbf{UCTR}  & \textbf{RS} 
            \\
            \midrule
            Online         & - & -  & - & -   \\
            LREF          & +0.023\%   & +0.209\%  & +0.120\%    & +1.016\%   \\
            \bottomrule
        \end{tabular}
    }
\end{table}
we observe that Our LREF frame shows demonstrates great improvements in both the UCVR and Relevance Satisfaction. (1) The great improvement in Relevance Satisfaction shows that more relevant products are displayed in the apps, enhancing the user experiences. (2)The UCTR improvement indicates that users are provided with more relevant products that match their intents leading to higher satisfaction click rates (3) The growth of UCVR indicates that presenting more relevant products leads to an increase in user clicks and purchases. These results underline that employing the LREF approach improves both user satisfaction and business efficiency.
\section{Conclusion}

This paper introduces a novel framework, which is called \textbf{L}LM-based \textbf{RE}levance \textbf{F}ramework (\textbf{LREF}), designed to enhance search relevance in e-commerce applications.  
Aiming to address the challenges of applying LLMs to query-product relevance classification tasks, LREF  offers a novel way to select high-quality data in large noisy human annotation datasets, involves an internal guide for task-specific reasoning, and applies a DPO-based De-biasing approach to mitigate LLM's optimistic bias in relevance decisions. We conduct a series of offline experiments on large-scale real-world datasets and online A/B testing. The experiments show that our framework achieves significant improvements in both offline and online metrics. Finally, the model is deployed on a well-known e-commerce application and achieves significant results.



\begin{spacing}{1.2}
\bibliographystyle{ACM-Reference-Format}
\balance
\bibliography{sample-base}


\begin{thebibliography}{34}


\ifx \showCODEN    \undefined \def \showCODEN     #1{\unskip}     \fi
\ifx \showDOI      \undefined \def \showDOI       #1{#1}\fi
\ifx \showISBNx    \undefined \def \showISBNx     #1{\unskip}     \fi
\ifx \showISBNxiii \undefined \def \showISBNxiii  #1{\unskip}     \fi
\ifx \showISSN     \undefined \def \showISSN      #1{\unskip}     \fi
\ifx \showLCCN     \undefined \def \showLCCN      #1{\unskip}     \fi
\ifx \shownote     \undefined \def \shownote      #1{#1}          \fi
\ifx \showarticletitle \undefined \def \showarticletitle #1{#1}   \fi
\ifx \showURL      \undefined \def \showURL       {\relax}        \fi
\providecommand\bibfield[2]{#2}
\providecommand\bibinfo[2]{#2}
\providecommand\natexlab[1]{#1}
\providecommand\showeprint[2][]{arXiv:#2}

\bibitem[Aizawa(2003)]%
        {aizawa2003information}
\bibfield{author}{\bibinfo{person}{Akiko Aizawa}.} \bibinfo{year}{2003}\natexlab{}.
\newblock \showarticletitle{An information-theoretic perspective of tf--idf measures}.
\newblock \bibinfo{journal}{\emph{Information Processing \& Management}} \bibinfo{volume}{39}, \bibinfo{number}{1} (\bibinfo{year}{2003}), \bibinfo{pages}{45--65}.
\newblock


\bibitem[Cao et~al\mbox{.}(2023)]%
        {cao2023instruction}
\bibfield{author}{\bibinfo{person}{Yihan Cao}, \bibinfo{person}{Yanbin Kang}, \bibinfo{person}{Chi Wang}, {and} \bibinfo{person}{Lichao Sun}.} \bibinfo{year}{2023}\natexlab{}.
\newblock \showarticletitle{Instruction Mining: Instruction Data Selection for Tuning Large Language Models}.
\newblock \bibinfo{journal}{\emph{arXiv preprint arXiv:2307.06290}} (\bibinfo{year}{2023}).
\newblock


\bibitem[Chang et~al\mbox{.}(2024)]%
        {chang2024survey}
\bibfield{author}{\bibinfo{person}{Yupeng Chang}, \bibinfo{person}{Xu Wang}, \bibinfo{person}{Jindong Wang}, \bibinfo{person}{Yuan Wu}, \bibinfo{person}{Linyi Yang}, \bibinfo{person}{Kaijie Zhu}, \bibinfo{person}{Hao Chen}, \bibinfo{person}{Xiaoyuan Yi}, \bibinfo{person}{Cunxiang Wang}, \bibinfo{person}{Yidong Wang}, {et~al\mbox{.}}} \bibinfo{year}{2024}\natexlab{}.
\newblock \showarticletitle{A survey on evaluation of large language models}.
\newblock \bibinfo{journal}{\emph{ACM Transactions on Intelligent Systems and Technology}} \bibinfo{volume}{15}, \bibinfo{number}{3} (\bibinfo{year}{2024}), \bibinfo{pages}{1--45}.
\newblock


\bibitem[Choudhary et~al\mbox{.}(2024)]%
        {choudhary2024interpretable}
\bibfield{author}{\bibinfo{person}{Nurendra Choudhary}, \bibinfo{person}{Edward~W Huang}, \bibinfo{person}{Karthik Subbian}, {and} \bibinfo{person}{Chandan~K Reddy}.} \bibinfo{year}{2024}\natexlab{}.
\newblock \showarticletitle{An interpretable ensemble of graph and language models for improving search relevance in e-commerce}. In \bibinfo{booktitle}{\emph{Companion Proceedings of the ACM on Web Conference 2024}}. \bibinfo{pages}{206--215}.
\newblock


\bibitem[Chu et~al\mbox{.}(2023)]%
        {chu2023survey}
\bibfield{author}{\bibinfo{person}{Zheng Chu}, \bibinfo{person}{Jingchang Chen}, \bibinfo{person}{Qianglong Chen}, \bibinfo{person}{Weijiang Yu}, \bibinfo{person}{Tao He}, \bibinfo{person}{Haotian Wang}, \bibinfo{person}{Weihua Peng}, \bibinfo{person}{Ming Liu}, \bibinfo{person}{Bing Qin}, {and} \bibinfo{person}{Ting Liu}.} \bibinfo{year}{2023}\natexlab{}.
\newblock \showarticletitle{A survey of chain of thought reasoning: Advances, frontiers and future}.
\newblock \bibinfo{journal}{\emph{arXiv preprint arXiv:2309.15402}} (\bibinfo{year}{2023}).
\newblock


\bibitem[Devlin(2018)]%
        {devlin2018bert}
\bibfield{author}{\bibinfo{person}{Jacob Devlin}.} \bibinfo{year}{2018}\natexlab{}.
\newblock \showarticletitle{Bert: Pre-training of deep bidirectional transformers for language understanding}.
\newblock \bibinfo{journal}{\emph{arXiv preprint arXiv:1810.04805}} (\bibinfo{year}{2018}).
\newblock


\bibitem[Eigner and H{\"a}ndler(2024)]%
        {eigner2024determinants}
\bibfield{author}{\bibinfo{person}{Eva Eigner} {and} \bibinfo{person}{Thorsten H{\"a}ndler}.} \bibinfo{year}{2024}\natexlab{}.
\newblock \showarticletitle{Determinants of llm-assisted decision-making}.
\newblock \bibinfo{journal}{\emph{arXiv preprint arXiv:2402.17385}} (\bibinfo{year}{2024}).
\newblock


\bibitem[Ethayarajh et~al\mbox{.}(2024)]%
        {ethayarajh2024kto}
\bibfield{author}{\bibinfo{person}{Kawin Ethayarajh}, \bibinfo{person}{Winnie Xu}, \bibinfo{person}{Niklas Muennighoff}, \bibinfo{person}{Dan Jurafsky}, {and} \bibinfo{person}{Douwe Kiela}.} \bibinfo{year}{2024}\natexlab{}.
\newblock \showarticletitle{Kto: Model alignment as prospect theoretic optimization}.
\newblock \bibinfo{journal}{\emph{arXiv preprint arXiv:2402.01306}} (\bibinfo{year}{2024}).
\newblock


\bibitem[Floridi and Chiriatti(2020)]%
        {floridi2020gpt}
\bibfield{author}{\bibinfo{person}{Luciano Floridi} {and} \bibinfo{person}{Massimo Chiriatti}.} \bibinfo{year}{2020}\natexlab{}.
\newblock \showarticletitle{GPT-3: Its nature, scope, limits, and consequences}.
\newblock \bibinfo{journal}{\emph{Minds and Machines}}  \bibinfo{volume}{30} (\bibinfo{year}{2020}), \bibinfo{pages}{681--694}.
\newblock


\bibitem[Gao et~al\mbox{.}(2023)]%
        {gao2023llama}
\bibfield{author}{\bibinfo{person}{Peng Gao}, \bibinfo{person}{Jiaming Han}, \bibinfo{person}{Renrui Zhang}, \bibinfo{person}{Ziyi Lin}, \bibinfo{person}{Shijie Geng}, \bibinfo{person}{Aojun Zhou}, \bibinfo{person}{Wei Zhang}, \bibinfo{person}{Pan Lu}, \bibinfo{person}{Conghui He}, \bibinfo{person}{Xiangyu Yue}, {et~al\mbox{.}}} \bibinfo{year}{2023}\natexlab{}.
\newblock \showarticletitle{Llama-adapter v2: Parameter-efficient visual instruction model}.
\newblock \bibinfo{journal}{\emph{arXiv preprint arXiv:2304.15010}} (\bibinfo{year}{2023}).
\newblock


\bibitem[He et~al\mbox{.}(2020)]%
        {he2020deberta}
\bibfield{author}{\bibinfo{person}{Pengcheng He}, \bibinfo{person}{Xiaodong Liu}, \bibinfo{person}{Jianfeng Gao}, {and} \bibinfo{person}{Weizhu Chen}.} \bibinfo{year}{2020}\natexlab{}.
\newblock \showarticletitle{Deberta: Decoding-enhanced bert with disentangled attention}.
\newblock \bibinfo{journal}{\emph{arXiv preprint arXiv:2006.03654}} (\bibinfo{year}{2020}).
\newblock


\bibitem[Kaufmann et~al\mbox{.}(2023)]%
        {kaufmann2023survey}
\bibfield{author}{\bibinfo{person}{Timo Kaufmann}, \bibinfo{person}{Paul Weng}, \bibinfo{person}{Viktor Bengs}, {and} \bibinfo{person}{Eyke H{\"u}llermeier}.} \bibinfo{year}{2023}\natexlab{}.
\newblock \showarticletitle{A survey of reinforcement learning from human feedback}.
\newblock \bibinfo{journal}{\emph{arXiv preprint arXiv:2312.14925}} (\bibinfo{year}{2023}).
\newblock


\bibitem[Li et~al\mbox{.}(2023)]%
        {li2023quantity}
\bibfield{author}{\bibinfo{person}{Ming Li}, \bibinfo{person}{Yong Zhang}, \bibinfo{person}{Zhitao Li}, \bibinfo{person}{Jiuhai Chen}, \bibinfo{person}{Lichang Chen}, \bibinfo{person}{Ning Cheng}, \bibinfo{person}{Jianzong Wang}, \bibinfo{person}{Tianyi Zhou}, {and} \bibinfo{person}{Jing Xiao}.} \bibinfo{year}{2023}\natexlab{}.
\newblock \showarticletitle{From quantity to quality: Boosting llm performance with self-guided data selection for instruction tuning}.
\newblock \bibinfo{journal}{\emph{arXiv preprint arXiv:2308.12032}} (\bibinfo{year}{2023}).
\newblock


\bibitem[Liu et~al\mbox{.}(2024)]%
        {liu2024take}
\bibfield{author}{\bibinfo{person}{Ziche Liu}, \bibinfo{person}{Rui Ke}, \bibinfo{person}{Feng Jiang}, {and} \bibinfo{person}{Haizhou Li}.} \bibinfo{year}{2024}\natexlab{}.
\newblock \showarticletitle{Take the essence and discard the dross: A Rethinking on Data Selection for Fine-Tuning Large Language Models}.
\newblock \bibinfo{journal}{\emph{arXiv preprint arXiv:2406.14115}} (\bibinfo{year}{2024}).
\newblock


\bibitem[Liu et~al\mbox{.}(2022)]%
        {liu2022knowledge}
\bibfield{author}{\bibinfo{person}{Ziyang Liu}, \bibinfo{person}{Chaokun Wang}, \bibinfo{person}{Hao Feng}, \bibinfo{person}{Lingfei Wu}, {and} \bibinfo{person}{Liqun Yang}.} \bibinfo{year}{2022}\natexlab{}.
\newblock \showarticletitle{Knowledge distillation based contextual relevance matching for e-commerce product search}. In \bibinfo{booktitle}{\emph{Proceedings of the 2022 Conference on Empirical Methods in Natural Language Processing: Industry Track}}. \bibinfo{pages}{63--76}.
\newblock


\bibitem[Mehrdad et~al\mbox{.}(2024)]%
        {mehrdad2024large}
\bibfield{author}{\bibinfo{person}{Navid Mehrdad}, \bibinfo{person}{Hrushikesh Mohapatra}, \bibinfo{person}{Mossaab Bagdouri}, \bibinfo{person}{Prijith Chandran}, \bibinfo{person}{Alessandro Magnani}, \bibinfo{person}{Xunfan Cai}, \bibinfo{person}{Ajit Puthenputhussery}, \bibinfo{person}{Sachin Yadav}, \bibinfo{person}{Tony Lee}, \bibinfo{person}{ChengXiang Zhai}, {et~al\mbox{.}}} \bibinfo{year}{2024}\natexlab{}.
\newblock \showarticletitle{Large Language Models for Relevance Judgment in Product Search}.
\newblock \bibinfo{journal}{\emph{arXiv preprint arXiv:2406.00247}} (\bibinfo{year}{2024}).
\newblock


\bibitem[Mikolov et~al\mbox{.}(2013)]%
        {mikolov2013distributed}
\bibfield{author}{\bibinfo{person}{Tomas Mikolov}, \bibinfo{person}{Ilya Sutskever}, \bibinfo{person}{Kai Chen}, \bibinfo{person}{Greg~S Corrado}, {and} \bibinfo{person}{Jeff Dean}.} \bibinfo{year}{2013}\natexlab{}.
\newblock \showarticletitle{Distributed representations of words and phrases and their compositionality}.
\newblock \bibinfo{journal}{\emph{Advances in neural information processing systems}}  \bibinfo{volume}{26} (\bibinfo{year}{2013}).
\newblock


\bibitem[Naveed et~al\mbox{.}(2023)]%
        {naveed2023comprehensive}
\bibfield{author}{\bibinfo{person}{Humza Naveed}, \bibinfo{person}{Asad~Ullah Khan}, \bibinfo{person}{Shi Qiu}, \bibinfo{person}{Muhammad Saqib}, \bibinfo{person}{Saeed Anwar}, \bibinfo{person}{Muhammad Usman}, \bibinfo{person}{Naveed Akhtar}, \bibinfo{person}{Nick Barnes}, {and} \bibinfo{person}{Ajmal Mian}.} \bibinfo{year}{2023}\natexlab{}.
\newblock \showarticletitle{A comprehensive overview of large language models}.
\newblock \bibinfo{journal}{\emph{arXiv preprint arXiv:2307.06435}} (\bibinfo{year}{2023}).
\newblock


\bibitem[Pennington et~al\mbox{.}(2014)]%
        {pennington2014glove}
\bibfield{author}{\bibinfo{person}{Jeffrey Pennington}, \bibinfo{person}{Richard Socher}, {and} \bibinfo{person}{Christopher~D Manning}.} \bibinfo{year}{2014}\natexlab{}.
\newblock \showarticletitle{Glove: Global vectors for word representation}. In \bibinfo{booktitle}{\emph{Proceedings of the 2014 conference on empirical methods in natural language processing (EMNLP)}}. \bibinfo{pages}{1532--1543}.
\newblock


\bibitem[Rafailov et~al\mbox{.}(2024)]%
        {rafailov2024direct}
\bibfield{author}{\bibinfo{person}{Rafael Rafailov}, \bibinfo{person}{Archit Sharma}, \bibinfo{person}{Eric Mitchell}, \bibinfo{person}{Christopher~D Manning}, \bibinfo{person}{Stefano Ermon}, {and} \bibinfo{person}{Chelsea Finn}.} \bibinfo{year}{2024}\natexlab{}.
\newblock \showarticletitle{Direct preference optimization: Your language model is secretly a reward model}.
\newblock \bibinfo{journal}{\emph{Advances in Neural Information Processing Systems}}  \bibinfo{volume}{36} (\bibinfo{year}{2024}).
\newblock


\bibitem[Renze and Guven(2024)]%
        {renze2024self}
\bibfield{author}{\bibinfo{person}{Matthew Renze} {and} \bibinfo{person}{Erhan Guven}.} \bibinfo{year}{2024}\natexlab{}.
\newblock \showarticletitle{Self-Reflection in LLM Agents: Effects on Problem-Solving Performance}.
\newblock \bibinfo{journal}{\emph{arXiv preprint arXiv:2405.06682}} (\bibinfo{year}{2024}).
\newblock


\bibitem[Touvron et~al\mbox{.}(2023)]%
        {touvron2023llama}
\bibfield{author}{\bibinfo{person}{Hugo Touvron}, \bibinfo{person}{Thibaut Lavril}, \bibinfo{person}{Gautier Izacard}, \bibinfo{person}{Xavier Martinet}, \bibinfo{person}{Marie-Anne Lachaux}, \bibinfo{person}{Timoth{\'e}e Lacroix}, \bibinfo{person}{Baptiste Rozi{\`e}re}, \bibinfo{person}{Naman Goyal}, \bibinfo{person}{Eric Hambro}, \bibinfo{person}{Faisal Azhar}, {et~al\mbox{.}}} \bibinfo{year}{2023}\natexlab{}.
\newblock \showarticletitle{Llama: Open and efficient foundation language models}.
\newblock \bibinfo{journal}{\emph{arXiv preprint arXiv:2302.13971}} (\bibinfo{year}{2023}).
\newblock


\bibitem[Vaswani(2017)]%
        {vaswani2017attention}
\bibfield{author}{\bibinfo{person}{A Vaswani}.} \bibinfo{year}{2017}\natexlab{}.
\newblock \showarticletitle{Attention is all you need}.
\newblock \bibinfo{journal}{\emph{Advances in Neural Information Processing Systems}} (\bibinfo{year}{2017}).
\newblock


\bibitem[Villaflor et~al\mbox{.}(2022)]%
        {villaflor2022addressing}
\bibfield{author}{\bibinfo{person}{Adam~R Villaflor}, \bibinfo{person}{Zhe Huang}, \bibinfo{person}{Swapnil Pande}, \bibinfo{person}{John~M Dolan}, {and} \bibinfo{person}{Jeff Schneider}.} \bibinfo{year}{2022}\natexlab{}.
\newblock \showarticletitle{Addressing optimism bias in sequence modeling for reinforcement learning}. In \bibinfo{booktitle}{\emph{international conference on machine learning}}. PMLR, \bibinfo{pages}{22270--22283}.
\newblock


\bibitem[Wang et~al\mbox{.}(2024a)]%
        {wang2024utilizing}
\bibfield{author}{\bibinfo{person}{Jiajia Wang}, \bibinfo{person}{Jimmy~Xiangji Huang}, \bibinfo{person}{Xinhui Tu}, \bibinfo{person}{Junmei Wang}, \bibinfo{person}{Angela~Jennifer Huang}, \bibinfo{person}{Md~Tahmid~Rahman Laskar}, {and} \bibinfo{person}{Amran Bhuiyan}.} \bibinfo{year}{2024}\natexlab{a}.
\newblock \showarticletitle{Utilizing BERT for Information Retrieval: Survey, Applications, Resources, and Challenges}.
\newblock \bibinfo{journal}{\emph{Comput. Surveys}} \bibinfo{volume}{56}, \bibinfo{number}{7} (\bibinfo{year}{2024}), \bibinfo{pages}{1--33}.
\newblock


\bibitem[Wang et~al\mbox{.}(2024b)]%
        {wang2024survey}
\bibfield{author}{\bibinfo{person}{Jiahao Wang}, \bibinfo{person}{Bolin Zhang}, \bibinfo{person}{Qianlong Du}, \bibinfo{person}{Jiajun Zhang}, {and} \bibinfo{person}{Dianhui Chu}.} \bibinfo{year}{2024}\natexlab{b}.
\newblock \showarticletitle{A Survey on Data Selection for LLM Instruction Tuning}.
\newblock \bibinfo{journal}{\emph{arXiv preprint arXiv:2402.05123}} (\bibinfo{year}{2024}).
\newblock


\bibitem[Wang et~al\mbox{.}(2022)]%
        {wang2022self}
\bibfield{author}{\bibinfo{person}{Xuezhi Wang}, \bibinfo{person}{Jason Wei}, \bibinfo{person}{Dale Schuurmans}, \bibinfo{person}{Quoc Le}, \bibinfo{person}{Ed Chi}, \bibinfo{person}{Sharan Narang}, \bibinfo{person}{Aakanksha Chowdhery}, {and} \bibinfo{person}{Denny Zhou}.} \bibinfo{year}{2022}\natexlab{}.
\newblock \showarticletitle{Self-consistency improves chain of thought reasoning in language models}.
\newblock \bibinfo{journal}{\emph{arXiv preprint arXiv:2203.11171}} (\bibinfo{year}{2022}).
\newblock


\bibitem[Wei et~al\mbox{.}(2022)]%
        {wei2022chain}
\bibfield{author}{\bibinfo{person}{Jason Wei}, \bibinfo{person}{Xuezhi Wang}, \bibinfo{person}{Dale Schuurmans}, \bibinfo{person}{Maarten Bosma}, \bibinfo{person}{Fei Xia}, \bibinfo{person}{Ed Chi}, \bibinfo{person}{Quoc~V Le}, \bibinfo{person}{Denny Zhou}, {et~al\mbox{.}}} \bibinfo{year}{2022}\natexlab{}.
\newblock \showarticletitle{Chain-of-thought prompting elicits reasoning in large language models}.
\newblock \bibinfo{journal}{\emph{Advances in neural information processing systems}}  \bibinfo{volume}{35} (\bibinfo{year}{2022}), \bibinfo{pages}{24824--24837}.
\newblock


\bibitem[Wiseman and Rush(2016)]%
        {wiseman2016sequence}
\bibfield{author}{\bibinfo{person}{Sam Wiseman} {and} \bibinfo{person}{Alexander~M Rush}.} \bibinfo{year}{2016}\natexlab{}.
\newblock \showarticletitle{Sequence-to-sequence learning as beam-search optimization}.
\newblock \bibinfo{journal}{\emph{arXiv preprint arXiv:1606.02960}} (\bibinfo{year}{2016}).
\newblock


\bibitem[Xia et~al\mbox{.}(2024)]%
        {xia2024less}
\bibfield{author}{\bibinfo{person}{Mengzhou Xia}, \bibinfo{person}{Sadhika Malladi}, \bibinfo{person}{Suchin Gururangan}, \bibinfo{person}{Sanjeev Arora}, {and} \bibinfo{person}{Danqi Chen}.} \bibinfo{year}{2024}\natexlab{}.
\newblock \showarticletitle{Less: Selecting influential data for targeted instruction tuning}.
\newblock \bibinfo{journal}{\emph{arXiv preprint arXiv:2402.04333}} (\bibinfo{year}{2024}).
\newblock


\bibitem[Yan et~al\mbox{.}(2018)]%
        {yan2018beyond}
\bibfield{author}{\bibinfo{person}{Su Yan}, \bibinfo{person}{Wei Lin}, \bibinfo{person}{Tianshu Wu}, \bibinfo{person}{Daorui Xiao}, \bibinfo{person}{Xu Zheng}, \bibinfo{person}{Bo Wu}, {and} \bibinfo{person}{Kaipeng Liu}.} \bibinfo{year}{2018}\natexlab{}.
\newblock \showarticletitle{Beyond keywords and relevance: a personalized ad retrieval framework in e-commerce sponsored search}. In \bibinfo{booktitle}{\emph{Proceedings of the 2018 World Wide Web Conference}}. \bibinfo{pages}{1919--1928}.
\newblock


\bibitem[Yuan et~al\mbox{.}(2024)]%
        {yuan2024semi}
\bibfield{author}{\bibinfo{person}{Chunyuan Yuan}, \bibinfo{person}{Ming Pang}, \bibinfo{person}{Zheng Fang}, \bibinfo{person}{Xue Jiang}, \bibinfo{person}{Changping Peng}, {and} \bibinfo{person}{Zhangang Lin}.} \bibinfo{year}{2024}\natexlab{}.
\newblock \showarticletitle{A Semi-supervised Multi-channel Graph Convolutional Network for Query Classification in E-commerce}. In \bibinfo{booktitle}{\emph{Companion Proceedings of the ACM on Web Conference 2024}}. \bibinfo{pages}{56--64}.
\newblock


\bibitem[Yuan et~al\mbox{.}(2023)]%
        {yuan2023multi}
\bibfield{author}{\bibinfo{person}{Chunyuan Yuan}, \bibinfo{person}{Yiming Qiu}, \bibinfo{person}{Mingming Li}, \bibinfo{person}{Haiqing Hu}, \bibinfo{person}{Songlin Wang}, {and} \bibinfo{person}{Sulong Xu}.} \bibinfo{year}{2023}\natexlab{}.
\newblock \showarticletitle{A multi-granularity matching attention network for query intent classification in e-commerce retrieval}. In \bibinfo{booktitle}{\emph{Companion Proceedings of the ACM Web Conference 2023}}. \bibinfo{pages}{416--420}.
\newblock


\bibitem[Zhang et~al\mbox{.}(2022)]%
        {zhang2022automatic}
\bibfield{author}{\bibinfo{person}{Zhuosheng Zhang}, \bibinfo{person}{Aston Zhang}, \bibinfo{person}{Mu Li}, {and} \bibinfo{person}{Alex Smola}.} \bibinfo{year}{2022}\natexlab{}.
\newblock \showarticletitle{Automatic chain of thought prompting in large language models}.
\newblock \bibinfo{journal}{\emph{arXiv preprint arXiv:2210.03493}} (\bibinfo{year}{2022}).
\newblock


\end{thebibliography}
\end{spacing}

\end{document}